\documentstyle[12pt,epsf]{article}
\renewcommand{\baselinestretch}{1.0}

\newcommand{\be}{\begin{equation}}
\newcommand{\ee}{\end{equation}}
\begin{document}
\topmargin 0pt
\oddsidemargin=-0.4truecm
\evensidemargin=-0.4truecm
\renewcommand{\thefootnote}{\fnsymbol{footnote}}

\newpage
\setcounter{page}{1}
\begin{titlepage}
\vspace*{-2.0cm}
\begin{flushright}
\vspace*{-0.2cm}

\end{flushright}
\vspace*{0.5cm}

\begin{center}
{\Large \bf The MSW effect and Matter Effects in Neutrino Oscillations  
\footnote{Talk  given at  the Nobel Symposium 129, ``Neutrino Physics'',  
Haga Slott, August 19 - 24, 2004.}
}
\vspace{0.5cm}

{A. Yu. Smirnov$^{2,3}$\\

\vspace*{0.2cm}
{\em (2) The Abdus Salam International Centre for Theoretical Physics,
I-34100 Trieste, Italy }\\
{\em (3) Institute for Nuclear Research of Russian Academy
of Sciences, Moscow 117312, Russia}
}
\end{center}

\begin{abstract}

The MSW  (Mikheyev-Smirnov-Wolfenstein) 
effect is the adiabatic or partially adiabatic neutrino 
flavor conversion in medium with varying density.  
The main notions related to the effect, its dynamics and physical 
picture are reviewed. The large mixing MSW effect is realized inside the Sun 
providing the solution  of the  solar neutrino problem.
The small mixing MSW effect driven by the  1-3 mixing can 
be realized for the supernova (SN) neutrinos. Inside the collapsing stars  
new elements of the MSW dynamics may show up:   
the non-oscillatory transition, non-adiabatic conversion, 
time dependent adiabaticity violation induced  by  shock waves.
Effects of the resonance enhancement and the parametric enhancement of 
oscillations can be realized for the atmospheric 
and accelerator neutrinos in the Earth. 
Precise results for neutrino oscillations in the low 
density medium with arbitrary  density profile  are  presented  and the 
attenuation  effect is described. The area of applications  is the solar 
and SN neutrinos inside the 
Earth, and the results  are crucial for the neutrino oscillation 
tomography.

\end{abstract}
\end{titlepage}
\renewcommand{\thefootnote}{\arabic{footnote}}
\setcounter{footnote}{0}
\renewcommand{\baselinestretch}{0.9}

\section{Introduction}

Variety of matter effects~\cite{w1} on propagation of mixed neutrinos 
\cite{pontosc,mns} depends on 
(i) channel of mixing -  the type of neutrino states involved:   
active, sterile, mass eigenstates, {\it etc.}; 
(ii) density profile: constant, monotonous, periodic,   
fluctuating;   
(iii) properties of medium: polarization, motion, chemical composition   
(thermal bath and  neutrino gases are special cases);   
(iv) presence of new non-standard neutrino interactions. 

Dynamics of the matter effects can be classified by  degrees of freedom 
involved. In the $2\nu$ case,  
an arbitrary neutrino state can be expressed in terms of the
eigenstates of the instantaneous Hamiltonian, $\nu_{1m}$ and  $\nu_{2m}$,  
as
\be
\nu (x) = \cos\theta_a \nu_{1m} + \sin \theta_a \nu_{2m} e^{i\Phi_m}~,
\label{exp}
\ee
where
\begin{itemize}

\item
$\theta_a = \theta_a (x)$ determines the  admixtures of  
eigenstates in $\nu (x)$;

\item
$\Phi_m(x)$ is the phase difference between the two eigenstates (phase
of oscillations):
\be
\Phi_m(x) = \int_{x_0}^x \Delta H dt',  
\label{phase}
\ee  
here $\Delta H \equiv H_{2m} - H_{1m}$ is the difference  of eigenvalues.  
Eq. (\ref{phase})  gives the adiabatic phase,  
an additional contributions  can be related to non-adiabaticity 
and topology. 

\item 
The mixing angle in matter, $\theta_m$, defined as   
$\langle \nu_e| \nu_{1m}\rangle \equiv \cos \theta_m $, 
$\langle \nu_e| \nu_{2m}\rangle \equiv \sin \theta_m $, {\it etc.}, 
determines the  flavor contents (or flavors) of the eigenstates. The angle     
is the function of matter density $n_e(x)$: $\theta_m  = \theta_m(n_e(x))$.

\end{itemize}

Besides these,  the effects of absorption and loss of the coherence 
can change the normalization of the eigenstates or 
suppress interference.  

Different processes are associated with these  
degrees of freedom. In particular, in pure form  
the oscillations~\cite{pontosc,mns,pontprob,w1,bar} 
are the effect of the monotonous phase difference increase 
$\Phi_m$, which occurs in the uniform medium. 
In contrast, the MSW effect~\cite{w1,ms1} is associated to the change 
of  flavors of the eigenstates (change of $\theta_m(x)$) in the nonuniform medium. 
In general, an interplay of different effects occurs 
which is induced by simultaneous operation of several degrees of freedom. 

In this paper I will discuss only few effects selected on the 
following ground: usual massive neutrinos with the   flavor 
mixing  and standard weak interactions; the   
effects in the Sun, collapsing stars and the Earth;   
the effects which are realized in Nature or have a good chance to be realized 
and established in our  future studies. 

\section{ The MSW effect.}

The MSW effect~\cite{w1,ms1} is the adiabatic or partially adiabatic neutrino
flavor conversion in medium with varying density. The flavor 
of neutrino state follows the density change. 
Here I review the main notions related to the effect   
and describe its  dynamics. 

\subsection{Main notions}

{\it 1. Refraction. ~} 
At low energies the   {\it  elastic forward scattering} only is relevant 
in most of applications 
~\cite{w1,paul}.  It can be described  by the potential, $V$, or  
equivalently,  the refraction index:
$n_{ref} - 1 = {V}/{p}$.   

The difference of potentials (for a single neutrino) in 
different spatial points $V = V(x)$ leads  to 
the ``neutrino optics'' phenomena  such as  reflection, complete 
internal reflection, 
banding of neutrino trajectories, focusing by the stars, {\it etc.}.   
The values of refraction index are very close to unit, {\it e.g.},:  
$n_{ref} - 1 = 10^{-20}$ inside the Earth, $10^{-18}$  inside the Sun, 
$10^{-6}$ in the neutron stars. So, the ``neutrino lenses'' should be 
of the astrophysical size. 

The difference of potentials for different neutrinos, 
$\nu_e$ and $\nu_a$: $V \equiv V_e - V_a$  influences evolution of mixed 
neutrinos (even for constant potentials). In usual matter the  difference  
is due to  the charged current scattering 
of  $\nu_e$ on electrons  ($\nu_e e \rightarrow \nu_e e$) \cite{w1}: 
\be
V \equiv V_e - V_a  = \sqrt{2} G_F n_e~,
\ee
where $G_F$ is the Fermi coupling constant 
and $n_e$ is the number density of electrons. 
It leads to appearance of additional phase  in the neutrino system: 
$\Delta \phi_{matter} \equiv (V_e - V_a) t$. 
The distance over  which this ``matter" phase  equals 
$2\pi$ determines the {\it refraction length}: 
\be 
l_0 \equiv \frac{2\pi}{V_e - V_a} = \frac{\sqrt{2}\pi} {G_F n_e}.  
\ee

\noindent
{\it 2. Eigenstates and mixing in matter.~} 
In the  presence of matter the Hamiltonian of system changes: 
$
H_0  \rightarrow H = H_0 + V,
$
where $H_0$ is the Hamiltonian in vacuum. 
Correspondingly, the eigenstates and the eigenvalues of $H$  
change: 
$\nu_1, \nu_2 \rightarrow \nu_{1m}, \nu_{2m}$, 
$m_1^2/2E,  m_2^2/ 2E  \rightarrow  H_{1m}, H_{2m}$. 
Here $\nu_1$ and  $\nu_2$ are the mass eigenstates 
with masses  $m_1$ and $m_2$. 

The mixing in matter  is defined with respect 
to the eigenstates $\nu_{1m}$ and $\nu_{2m}$. 
Similarly to the vacuum case,  the mixing angle in  matter, $\theta_m$,  
determines  relations between the eigenstates in matter and  the flavor states: 
$\nu_e \equiv \cos\theta_m \nu_{1m} + \sin \theta_m \nu_{2m}$, 
{\it etc.}. 
In matter, both the eigenstates and eigenvalues, and consequently, 
the mixing angle depend on matter density and neutrino energy. 
\begin{figure}[htb]
\hbox to \hsize{\hfil\epsfxsize=9.0cm\epsfbox{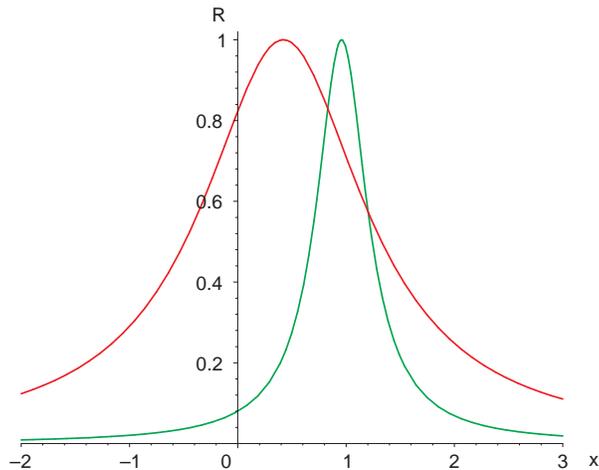}\hfil}
\caption{\small The dependence of the  effective mixing parameter 
$R \equiv \sin^2 2\theta_m$ 
on  $x \equiv l_{\nu}/l_0 \propto E n_e$ for two different values 
of the vacuum mixing: $\sin^2 2\theta = 0.825$ 
(black, red) which  is the LMA mixing and  $\tan^2 \theta = 0.08$ (grey, green).  
The semi-plane  $x < 0$ corresponds to the antineutrino channel. 
}
\label{fres}
\end{figure}

\noindent
{\it 3. Resonance.~} Dependence of the effective mixing 
parameter in matter, $R \equiv \sin^2 2\theta_m$, on ratio of the vacuum 
oscillation 
length, $l_{\nu} = 4\pi E/\Delta m^2$,  and the refraction length: 
$x \equiv {l_{\nu}}/{l_0} = 2 E V/ \Delta m^2 \propto E n_e$   
(fig.~\ref{fres}) has a resonance character~\cite{ms1}. At 
\be
l_{\nu} = l_0 \cos 2\theta~~~~~~~{\rm  (resonance~~ condition)} 
\label{res}
\ee
the mixing becomes maximal: $\sin^2 2\theta_m = 1$. 
For small $\theta$ the condition (\ref{res}) reads: 
\be
Vacuum~oscillation~length ~~\approx~~ Refraction~length. 
\label{res1}
\ee
That is, the eigenfrequency which characterizes a  system of mixed neutrinos, 
$1/l_{\nu}$,  coincides with the eigenfrequency of medium, $1/l_0$. 
 For large vacuum mixing (LMA has $\cos 2\theta = 0.4 $) there is a significant 
deviation from the equality (\ref{res1}). Large mixing corresponds 
to the case of strongly coupled system for 
which  the shift of frequencies occurs.

The resonance condition  (\ref{res}) determines the resonance density: 
\be 
n_e^{R} = n_0 \cos 2\theta, ~~~
n_0 \equiv  \frac{\Delta m^2}{2 \sqrt{2} E G_F}. 
\label{eq:resonance}
\ee
The  width of resonance on the half of height (in the density scale) 
is given by 
\be
2 \Delta n_e^R = 2 n_e^R \tan 2\theta = 2 n_0 \sin 2\theta.  
\label{width}
\ee
When the vacuum  mixing  approaches maximal value, $\theta = \pi/4$,   
the resonance shifts to zero density, $n_e^R \rightarrow 0$, and the width of  
resonance $\Delta n_e^R$  increases  converging to $n_0$. 

In medium with varying density, the resonance layer is determined by the 
interval in which the density changes  
from $n_e^R - \Delta n_e^R$  to $n_e^R + \Delta n_e^R$.

\noindent
{\it 4. Adiabaticity.~} Since in non-uniform medium the  
density changes on the way of neutrinos,  
$n_e = n_e(x)$,  the Hamiltonian of  system depends on 
time (distance):  $H = H(t)$. Therefore, 
(i) the mixing angle changes in course of  propagation: 
$\theta_m = \theta_m (n_e(x))$;
(ii) the eigenstates of instantaneous Hamiltonian, $\nu_{1m}$ and  
$\nu_{2m}$,  are no more the 
``eigenstates" of propagation,  and  the transitions $\nu_{1m} 
\leftrightarrow \nu_{2m}$ occur. 

If the density changes slowly,  
the system (mixed neutrinos) has time to adjust the change leading to 
the adiabatic evolution~\cite{w1,ms1,bet,mess}.  
The adiabaticity condition is \cite{mess}
\be
\gamma = \left| \frac{\dot{\theta}_m}{H_{2m}  - H_{1m}} \right| \ll 1. 
\label{adiab}
\ee
As follows from the evolution equation 
for the neutrino eigenstates \cite{ms1,mess}, 
$|\dot{\theta}_m|$ determines the energy of transition 
$\nu_{1m} \leftrightarrow \nu_{2m}$,  
and  $|H_{2m}  - H_{1m}|$ gives the energy gap between the levels. 
So, the condition (\ref{adiab}) means that 
the transitions $\nu_{1m} \leftrightarrow \nu_{2m}$ can be neglected 
and the eigenstates propagate independently 
($\theta_a = const$ in (\ref{exp})). 

If $\theta$ is small, 
the adiabaticity is  critical in the resonance. 
It takes the form \cite{ms1}
\be
\Delta r_R \geq l_R, 
\label{adiab2}
\ee
where $l_R = l_{\nu}/\sin 2\theta$ is the oscillation length in resonance, 
and $\Delta r_R = n_e^R \tan 2\theta / (dn_e/dr)_R$ 
is the spatial width of resonance layer. 
The adiabaticity condition has been considered outside the 
resonance and in the non-resonance channel in \cite{ssb}. 
In the case of large vacuum  mixing the point of maximal adiabaticity 
violation,  $n_e^{ad}$ ~\cite{lisi},  is shifted 
to densities larger than the resonance one: 
$n_e^{ad} \rightarrow n_0 > n^R$.\\

\subsection{Dynamics of the  MSW effect.}

{\it 1. Dynamical features~} of the effect in the adiabatic case  
can be summarized as follows.  

\begin{itemize}
 
\item
The flavors of  eigenstates change according to  density change;  
the flavors are determined by $\theta_m(x) = \theta_m(n_e(x))$. 

\item
The admixtures of the eigenstates in a propagating neutrino state do not change 
due to the adiabaticity; there is no $\nu_{1m} \leftrightarrow \nu_{2m}$ transitions. 
The  admixtures are given by the  mixing in  production point, $\theta_m^0$. 

\item
The phase difference $\Phi_m(x)$  (\ref{phase}) increases leading to 
oscillations.

\end{itemize}

Two degrees of freedom are operative: the phase $\Phi_m$ and 
the flavor, $\theta_m$. 
The MSW effect is  driven by the  change of  flavors of the  neutrino 
eigenstates in matter with varying density.  
The change of phase  produces the  oscillation effect on top of the 
adiabatic conversion.

\begin{figure}[htb]
\hbox to \hsize{\hfil\epsfxsize=9.0cm\epsfbox{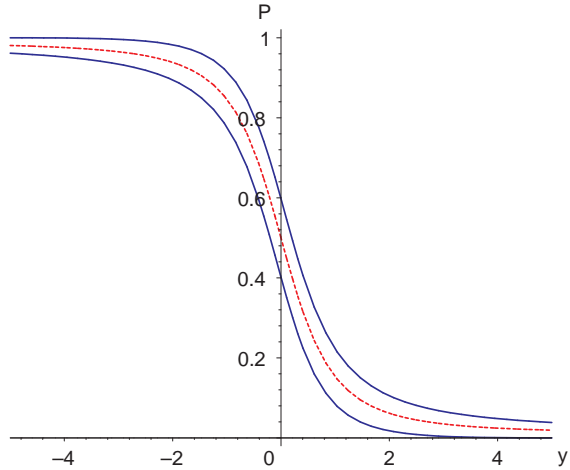}\hfil}
\caption{\small~~Dependences of the average  probability (dashed line) and the depth 
of oscillations ($P^{max}$, $P^{min}$ - solid lines)  
on $y$  for  $y_0 = - 5$.  The resonance layer corresponds to $y = 0$ 
and  the resonance layer is given by the interval $y = -1 \div  1$. 
For $\tan^2 \theta = 0.4$ (large mixing MSW solution) 
the evolution stops at $y_f = 0.47$. 
}
\label{msw}
\end{figure}

\noindent
{\it 2. Spatial picture of the MSW effect.~}  
In fig.~\ref{msw} shown are  dependences of the average 
probability,  $\bar{P}$, and depth 
of oscillations given by  $P^{max} - P^{min}$,  on  
\be
y \equiv \frac{n_e^R - n_e}{\Delta n_e^R}, 
\label{y-var}
\ee
the distance (in the density scale) from the resonance density  
in the units of  the width of resonance layer~\cite{ms1}. 
In terms of $n$ the 
conversion pattern depends on the initial value $n_0$ only which 
reflects {\it universality} of the adiabatic evolution. 
The probability is the oscillatory function which 
is inscribed into the band shown by the solid lines. 
There is no explicit dependence on the vacuum  angle $\theta$. 
With decrease of  $y_0$, the oscillation band  becomes narrower 
approaching the line of {\it  non-oscillatory conversion}. 
For zero final density:
$y_f = 1/\tan 2\theta.$ 
The smaller the mixing (and therefore,  the larger final $y_f$) the 
stronger 
transition. 

\noindent
{\it 3. Adiabaticity violation.~} If density changes rapidly, 
so that the condition  (\ref{adiab}) is not satisfied, 
the transitions $\nu_{1m} \leftrightarrow \nu_{2m}$ become efficient
~\cite{ms1,hax}. 
Therefore  admixtures of the eigenstates in a given propagating state 
change: $\theta_a = \theta_a(x)$.  Now all three degrees of freedom 
- phase, flavor, and admixture -  become operative.  
Typically, adiabaticity breaking leads to weakening of the  flavor  
transition and enhancement of oscillations.

\noindent
{\it 4. Graphic representation~ }\cite{ms4} is based on
\begin{figure}[htb]
\hbox to 
\hsize{\hfil\epsfxsize=5.0cm\epsfbox{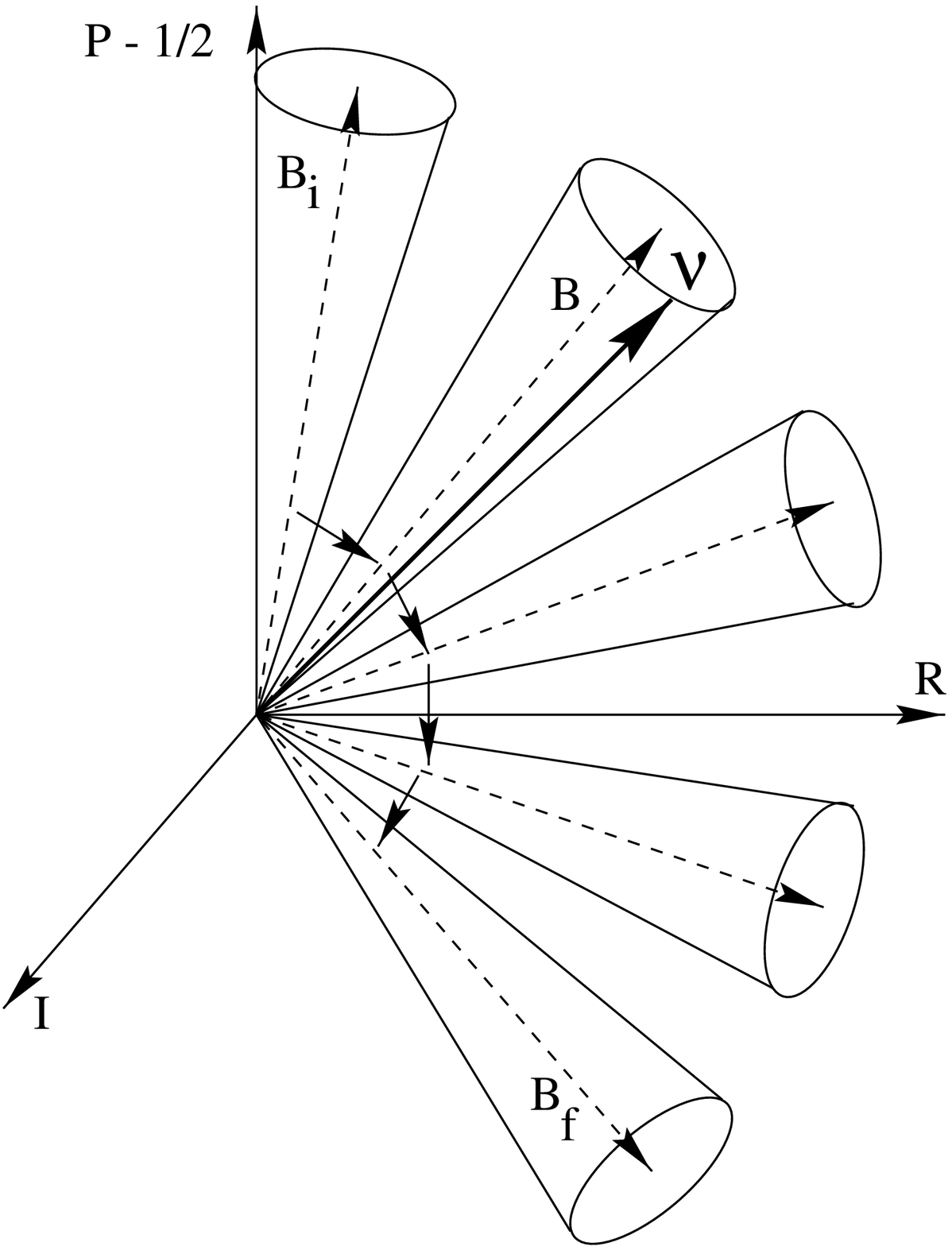}\hfil\epsfxsize=5.5cm\epsfbox{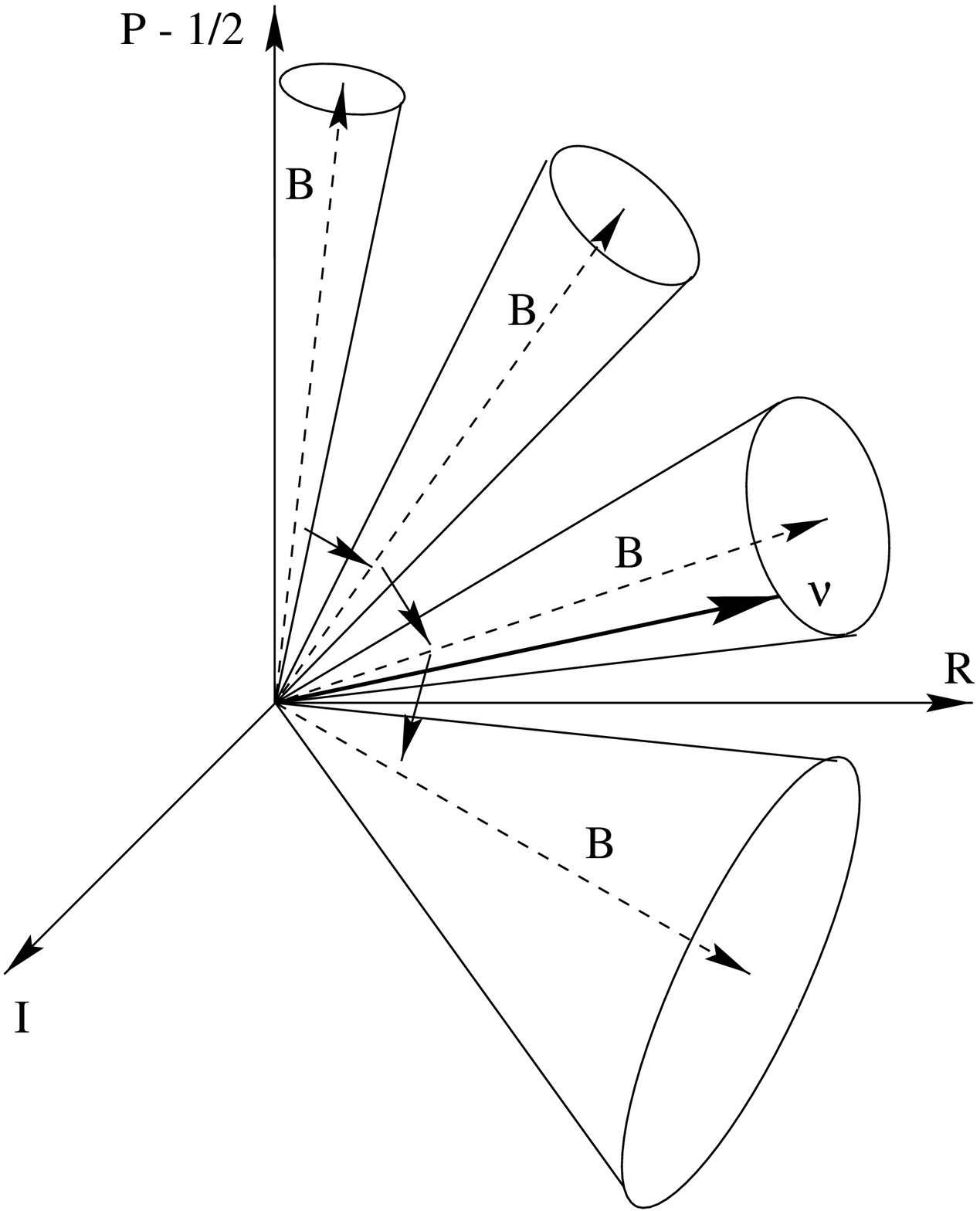}\hfil}
\caption{\small Graphic representation of the MSW effect. 
The direction of the 
cone axis is  determined by $2\theta_m$, the cone angle is given by 
$\theta_a$ and the  position of the neutrino vector on the surface of the 
cone is fixed by the phase $\Phi_m$. 
In the adiabatic case  (left panel) the direction of the axis flips 
according  to  $\theta_m{n_e}$ change, and the cone angle is unchanged. In 
the non-adiabatic case (right panel) also the cone angle changes.}
\label{graphmsw}
\end{figure}
the analogy of the neutrino evolution with the behavior of
spin of the electron in the magnetic field. 
The neutrino evolution equation  can be written as
\begin{equation}
\frac{d \vec{\nu}}{d t} = \left(\vec{B} \times \vec{\nu} \right)~, 
\end{equation}  
where the  neutrino vector of length 1/2 (equivalent of spin) is  
\begin{equation}
\vec{\nu} \equiv (R, I, P-1/2) = \left( {\rm Re} \nu_{e}^{\dagger} \nu_{\mu}, ~~
{\rm Im} \nu_{e}^{\dagger} \nu_{\mu},~~
\nu_{e}^{\dagger} \nu_{e} - 1/2 \right) , 
\label{graphnu}
\end{equation}
$\nu_{\alpha} = \nu_{\alpha}(x)$, ($\alpha =  \mu, e$) are the neutrino wave 
functions, 
and $P$ is the survival probability 
\footnote{The elements of this vector are nothing but  components of the density
matrix.}. The vector of ``magnetic field'' equals  
\begin{equation}
\vec{B} \equiv \frac{2 \pi}{l_m} (\sin 2 \theta_m,~~ 0 ,~~ \cos 2
\theta_m)~, 
\label{axisb}
\end{equation}
where $l_m = 2\pi/\Delta H$ is the oscillation length in matter. 
The vector $\vec{\nu}$  moves on the surface of the cone with axis $\vec{B}$
according to increase of the oscillation phase, $\Phi_m$.

\section{Realizations of the MSW effect}

General conditions for  the MSW conversion are: 
(i) slow enough density change;  (ii) crossing the resonance layer; 
(iii) large enough  matter width (minimal width condition) \cite{minw}. 
These conditions are satisfied for the solar neutrinos 
inside the Sun,  for supernova neutrinos  inside collapsing stars
and can be satisfied for neutrinos in the Early Universe.  


\subsection{Solar Neutrinos. Large Angle MSW solution}

The large mixing MSW conversion provides the solution of the solar neutrino 
problem~\cite{MD}. The best fit values of the oscillation parameters from  
combined analysis of the solar 
and KamLAND data (in assumption of the CPT invariance) are~\cite{kam}:   
\be
\Delta m^2 =  7.9 \cdot 10^{-5}~ {\rm eV}^2, ~~~~ \tan^2 \theta = 0.40. 
\label{paramo}
\ee 
Analysis of the solar neutrino data alone 
leads to smaller mass split,  
$\Delta m^2 =  6.3 \cdot 10^{-5} {\rm eV}^2$ which agrees with 
(\ref{paramo}) 
within $1\sigma$. 

\noindent
{\it 1.  Physical picture.~}
According to LMA,  inside the Sun the initially 
produced electron neutrinos 
undergo the highly  adiabatic conversion: $\nu_e \rightarrow \cos\theta_m^0 \nu_1 
+ \sin\theta_m^0 
\nu_2$, where $\theta_m^0$ is the mixing angle in the production point.  
On the way from the central parts of the Sun the coherence of  
neutrino state is lost after several hundreds oscillation 
lengths~\cite{HSsol}, and  incoherent fluxes 
of the mass states $\nu_1$ and $\nu_2$  arrive
at the surface of the Earth. In the matter of the Earth $\nu_1$ and $\nu_2$ 
oscillate partially regenerating the $\nu_e$-flux. 
The averaged survival probability can be written as 
\be
P = \sin^2 \theta +  \cos^2 \theta_{12}^{m0} \cos 2\theta_{12} - 
\cos 2\theta_{12}^{m0} f_{reg}, 
\label{surv}
\ee
where the first term corresponds to the non-oscillatory transition (dominates  
at the high energies), the second term is the contribution from the 
averaged oscillations which increases with decrease of  energy,  and the 
third term is the regeneration effect $f_{reg}$. 
At low energies $P$ reduces to the vacuum 
oscillation probability with very small matter corrections.  

\noindent
{\it 2. Status of LMA. ~ }
The solution provides a very good global fit of the solar neutrino data. 
There  is no statistically significant deviation from description given 
by the standard solar model (SSM)~\cite{john} and  the LMA solution. 

The key observation which testifies for the 
MSW (matter) effect in the Sun  is stronger than 1/2 
suppression of the signals at SK and SNO. 
The $\nu_e$-survival probability extracted from  
the CC/NC ratio at SNO  is 
\be 
\langle P_{ee} \rangle = 0.31^{+0.12}_{-0.08}    ~~~~(3\sigma),  
\label{eeration}
\ee
that is, $\langle P_{ee} \rangle < 0.50$ at $5\sigma$, 
whereas  the vacuum  $2\nu$ oscillations can produce $\langle P_{ee} \rangle \geq 0.5$.

Observations of two other signatures of the solution  
(i) the upturn of the spectrum at low energies ($E < 7 - 8$ MeV),  
(ii) the day-night asymmetry of signals with larger flux 
during the night are the main objectives of the forthcoming studies.


No viable alternative to the LMA solution exists and 
possible effects beyond LMA are substantially restricted already now.  

There is a very good agreement of the results from  solar neutrinos and  KamLAND 
which implies the CPT conservation. 
Furthermore,  it shows correctness of theory of both the vacuum 
oscillations and conversion in matter. 

\noindent
{\it 3.  Testing the MSW effect in the Sun. ~}
Important way to test the  effect is to introduce
the free parameter, $a_{MSW}$,  in the the matter potential
\be
V \rightarrow a_{MSW} V, 
\ee
and to determine $a_{MSW}$ from the data~\cite{lisiev}. 
The global analysis of the solar 
and reactor (KamLAND + CHOOZ) results with $\Delta m_{21}^2$, $\sin^2 
\theta_{12}$ and 
$a_{MSW}$ being unconstrained gives~\cite{lisiev} 
$a_{MSW} = 0.8 - 3.0$ (95\% C.L.) with 
the best fit value $a_{MSW} = 1.6$; zero value of $a_{MSW}$ is excluded at 
$6\sigma$.  

\noindent
{\it 4. Precision measurements in solar neutrinos.~}
Identification of the LMA solution opens new possibilities in~\cite{HLS} 
(i) precise description of the LMA conversion both inside the Sun and 
in the Earth taking into account various corrections; 
(ii) estimation of accuracy of approximation made; 
(iii) obtaining the accurate {\it analytic} expressions for probabilities 
and 
observables as functions of the oscillation parameters.  
There are three small quantities which allow for a  very precise expansions. 

(1). Smallness of the adiabaticity parameter 
\be
\gamma(x) = \frac{l_m}{4 \pi h(x)} \sim 10^{-3} - 10^{-4}, 
\label{adi}
\ee
where $h$ is the height of solar density profile,   
allows to use the adiabatic perturbation theory.  The non-adiabatic
corrections to the averaged survival probability are of the order 
$\gamma^2$~\cite{HLS}. 

(2). Smallness 
of the ratio $\Delta r_{prod}/h$, where $\Delta r_{prod}$ is the size 
of the neutrino production region, allows one to  
make the averaging of the survival probability 
over the neutrino production region in the analytic form~\cite{HLS}.    

(3). Smallness of the parameter 
\be
\epsilon (r) = \frac{2 E V_E}{\Delta m_{12}^2}  \leq 0.02 - 0.04, 
\label{earth}
\ee
where $V_E$ is the matter potential in the Earth, allows one  
to develop a very precise  perturbation theory for the neutrino 
oscillations 
inside the Earth~\cite{HLS}.

\subsection{MSW effect and Supernova neutrinos}

\begin{figure}[htb]
\hbox to \hsize{\hfil\epsfxsize=8.5cm\epsfbox{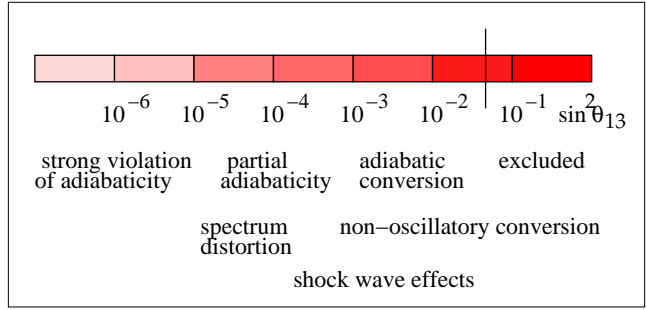}\hfil}
\caption{\small Scales of the 1-3 mixing probed with 
supernova neutrinos. Indicated are the regions of various effects in 
neutrino propagation. }
\label{tt}
\end{figure}

In supernovae one expects new elements of the MSW dynamics. The SN neutrinos  
probe whole $3\nu$ level crossing scheme,  
and  the effects of both resonances 
(due to $\Delta m^2_{12}$ and $\Delta m^2_{13}$)  should show up. 
Various effects associated to the 1-3 mixing 
can be realized, depending on value of $\theta_{13}$ (fig.~\ref{tt}). 
As follows from fig.~\ref{tt},  the SN neutrinos are sensitive to 
$\sin^2\theta_{13}$ as small as $10^{-5}$. 
Studies of the SN neutrinos   will also give information on 
the type of  mass hierarchy~\cite{DS,nun,barg,snsato,raffelt}.

\noindent
{\it 1. The small mixing MSW conversion. ~}
This can be realized due to the 1-3 mixing and the ``atmospheric''
mass split $\Delta m^2_{13}$. 

\noindent
{\it 2. The non-oscillatory adiabatic conversion ~}\cite{ms1} 
is expected for $\sin^2\theta_{13} > 10^{-3}$.  
The density in the production point is extremely large: $n(0) \gg n^R$, 
and therefore  the mixing in 
the initial state is strongly suppressed. So,  the neutrino state 
coincides with the eigenstate  in matter: $\nu_e = \nu_{2m}$. Due to  
adiabaticity  (no $\nu_{im} \leftrightarrow \nu_{jm}$ transition) the 
neutrino state coincides with this 
eigenstate  during whole evolution:  $\nu(t) =  \nu_{2m}$. 
The  interference is negligible since simply there is no second component to 
interfere with,  
and consequently, oscillations are absent. The flavor of the  neutrino 
state changes as the flavor of the eigenstate $\nu_{2m}$ and the latter 
follows the density change. 

\noindent
{\it 3. Adiabaticity violation~ } 
occurs if the 1-3 mixing is small $\sin^2\theta_{13} < 10^{-3}$.  

The shock wave can reach the region
of the neutrino conversion, $\rho \sim 10^{4}$ g/cc,
after $t_s = (3 - 5)$ s from the bounce 
(beginning of the $\nu-$ burst)~\cite{SF}.
Changing suddenly the density profile and therefore breaking the  
adiabaticity, 
the shock wave front influences the conversion in
the resonance characterized by  $\Delta m^2_{13}$ and $\sin^2 \theta_{13}$,
if $\sin^2 \theta_{13}  > 10^{-4}$.

The following shock wave effects can  be seen  
in neutrinos (antineutrinos)
for normal (inverted) hierarchy: 
(1) change of the total number of events in time \cite{SF};
(2) wave of softening of the spectrum which propagates
in the energy scale  from  low energies to high energies
\cite{Tak};
(3) delayed Earth matter effect
in the ``wrong" channel ({\it e.g.}, 
in neutrino channel for normal mass  hierarchy)
\cite{DS}.
Modification of the density profile by the shock wave leads to
appearance of additional resonances below the front~\cite{lisis}.  
Reverse shock produces a ``double dip'' time feature in the average 
neutrino energy~\cite{munich,raffelt}. 
Monitoring the shock wave with neutrinos 
can shed some light on the mechanism of explosion.

\noindent
{\it 4. Neutrinos from SN1987A.~} 
After confirmation of the LMA MSW solution we can definitely say that 
some effect of  flavor conversion has already been observed  in 1987.

In the case of  normal mass hierarchy the adiabatic 
$\bar{\nu}_e \rightarrow \bar \nu_1$
and   $\bar{\nu}_{\mu, \tau} \rightarrow \bar \nu_2$ transitions
occurred inside the star,  and then $\nu_1$ and $\nu_2$ oscillated inside
the Earth~\cite{ssb,DS}. In terms of the original fluxes of the  electron,
and muon antineutrinos,  $F^0(\bar \nu_e)$ and $F^0(\bar \nu_{\mu})$,  
the $\bar \nu_e-$ flux at the detector can be written as
\begin{equation}
F(\bar \nu_e) = F^0(\bar \nu_e) + \bar{p} \Delta F^0,
\label{flu}
\end{equation}
where $\Delta F^0 \equiv F(\bar \nu_{\mu}) -  F(\bar \nu_e)$,
and $\bar{p}$ is the {\it permutation} factor which can be calculated
precisely: $\bar{p} = \sin^2 \theta_{12} - \bar{f}_{reg}$.  
Due to difference in the distances traveled by neutrinos to
Kamiokande, IMB and Baksan detectors inside the Earth, 
the $\bar{\nu}_e$ regeneration factors  $\bar{f}_{reg}$ and therefore $\bar{p}$ 
differ for these detectors.
This can partially explain the difference of  the
Kamiokande and  IMB energy spectra of events~\cite{LS04}. 

One must take into account the conversion effects in analysis of
SN1987A~\cite{LS04} as well as future  supernova  neutrino data. 
The conversion can lead to  increase 
of the average energy of the observed events by 
(30 - 40)\%.  Inversely,  not taking into account
the conversion effect produces  errors 
in determination of the average energy of
the original  $\bar{\nu}_e$ spectrum up to  40 - 50 \% 
in Kamiokande, and  factor of 2 in IMB.

For the inverted mass hierarchy and  $\sin^2 \theta_{13} > 10^{-4}$
one would get nearly complete  permutation,
$\bar{p} \approx 1$, and therefore a harder $\bar\nu_e$ spectrum,
as well as an absence of the Earth matter effect. This is disfavored by  
the data~\cite{ssb,nun},  see however \cite{barg}.  
%

\section{Matter effects in Neutrino Oscillations}

Pure oscillation effect can be realized in the uniform medium.  
Mixing 
is constant, $\theta_m (E, n) = const$., and  therefore

\begin{itemize}

\item
the flavors of  eigenstates do not change;  

\item
the admixtures of  eigenstates do not change;   
there is no $\nu_{1m} \leftrightarrow \nu_{2m}$ transitions:  
$\nu_{1m}$ and  $\nu_{2m}$ are the eigenstates of propagation;  

\item
monotonous increase of $\Phi_m$ - the phase difference between the 
eigenstates occurs. 

\end{itemize}

Only one degree of freedom operates -  the phase $\Phi_m$  and all others 
are frozen. This is  similar to what happens in vacuum.  
The depht and length of oscillations  
are determined by the mixing and energy splitting in matter: 
$\sin^2 2\theta_m$, $l_{m} = 2\pi/(H_{2m} - H_{1m})$. 

The oscillations are  realized in the Earth  
which can be considered as  the multi-layer medium with 
nearly constant density in 
each layer.  Variety of possibilities exists depending on the 
neutrino trajectory 
(zenith angle), neutrino energy and channel of oscillations.

\subsection{Resonance enhancement of oscillations}

\begin{figure}[htb]
\hbox to 
\hsize{\hfil\epsfxsize=7.5cm\epsfbox{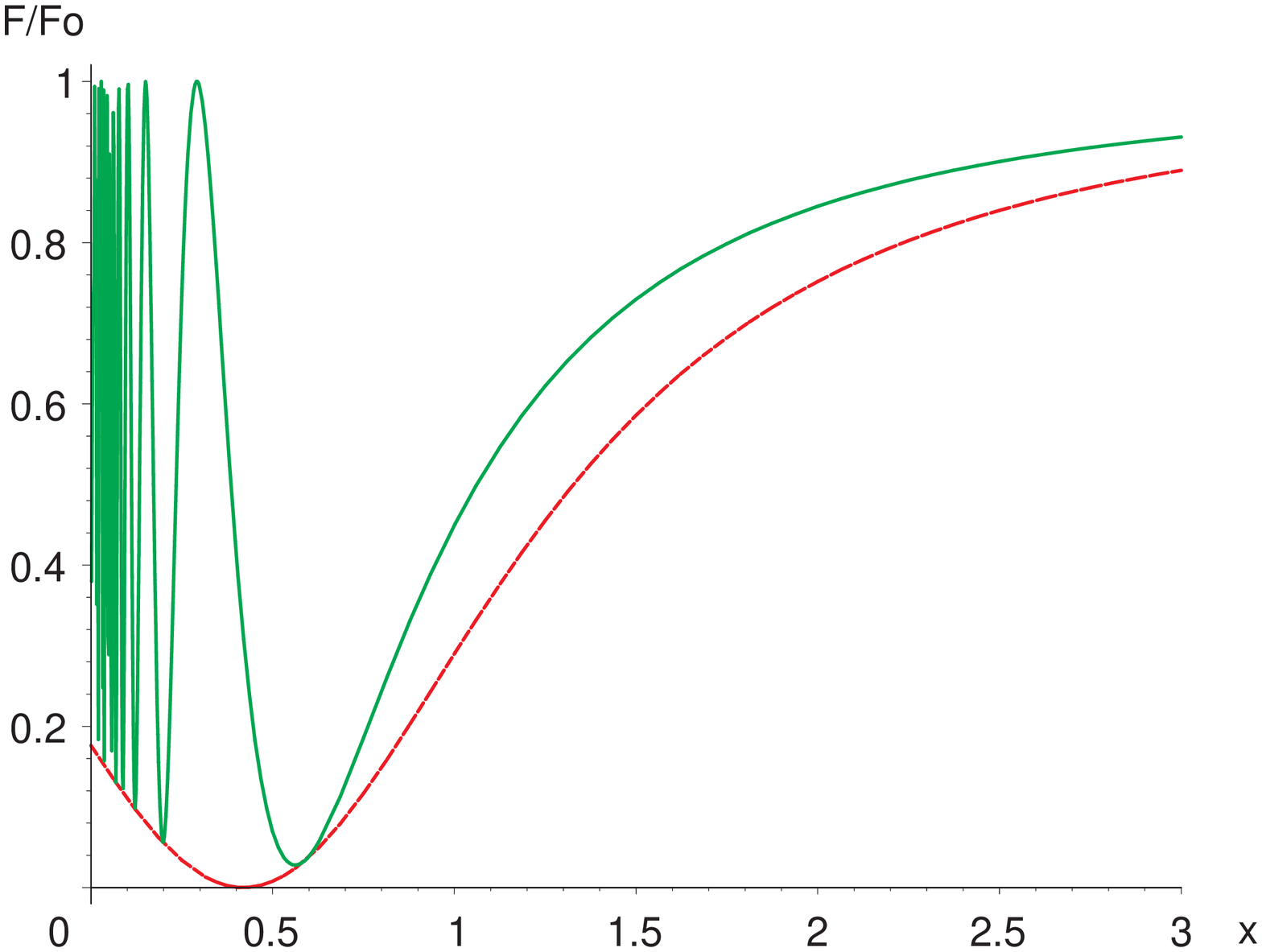}\hfil\epsfxsize=7.5cm\epsfbox{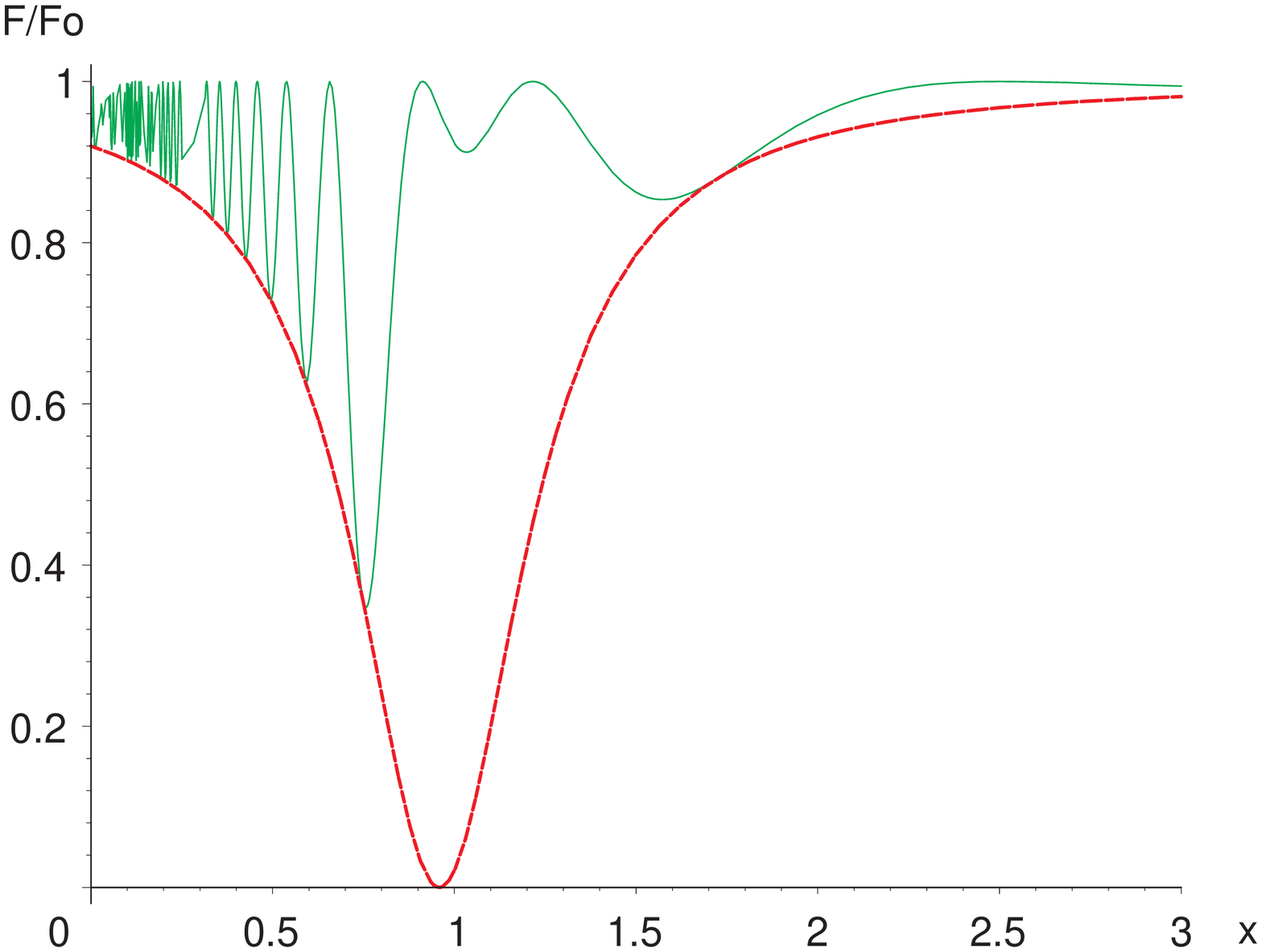}\hfil}
\caption{\small~Resonance enhancement of oscillations 
in matter with constant density.
Shown is the  dependence of  ratio of the final and original fluxes,  
(of {\it e.g.} $\nu_e$)  
$F/F_0$,  on energy ($x \equiv l_{\nu}/l_0\propto E$) for a thin layer, 
$L = l_0/\pi$, and $\sin^2 2\theta = 0.824$ 
(left panel), and thick  layer, $L =  10 l_0/ \pi$, 
and $\tan^2 \theta = 0.08$ (right panel). 
The oscillatory curves  are inscribed in to the resonance 
curve $(1 - \sin^2 2\theta_m)$. 
}
\label{resenh}
\end{figure}

For a given (constant) density, the length 
and depth of oscillations depend on the 
neutrino energy. This  leads to a characteristic 
distortion of the energy spectrum,   
$F(E)/F_0(E)$, where $F_0(E)$ and   $F(E)$  are,   
{\it e.g.} the spectra  of ({\it e.g.} $\nu_e$) neutrinos in the source 
and detector correspondingly (fig.~\ref{resenh}).  
The ratio $F(E)/F_0(E)$ given by the {\it survival probability} has an  
oscillatory dependence on the energy.  
At the resonance energy, $E_R$, the oscillations proceed with maximal 
depth;  they are enhanced in the resonance range~\cite{ms1}:  
$E = E_R \pm \Delta E_R,$  where 
$\Delta E_R = \tan 2\theta  E_R = \sin 2\theta  E_R^0$  
and  $E_R^0 = \Delta m^2/2\sqrt{2} G_F n_e$.  

The effect is realized for the high energy neutrinos ($E \sim$ GeV)  
in the mantle of  the 
Earth: constant density is a good first approximation. 
The effect is expected to be seen in the accelerator 
long baseline experiments~\cite{lind}, 
and future high statistics atmospheric neutrino studies~\cite{ADLS,ruiz}.

\subsection{Parametric enhancement of oscillations}

This  enhancement is related to  certain condition for the 
phase of oscillations \cite{ETC,Akh}. 
It provides another way of getting  strong transition: 
no matter enhancement or resonance conversion are needed.  No large or 
maximal mixings in vacuum or matter are  required.  

The simplest case which can be realized in Nature 
is neutrinos in the castle  wall profile. The latter consists 
of the alternate layers with two different densities~\cite{Akh,parliu,Pet}.  
Let   $\Phi_1$, $\Phi_2$ and $\theta_{1}^m$, $\theta_{2}^m$ be  
the phases and mixing angles in the layers 1 and 2.   
\begin{figure}[htb]
\hbox to
\hsize{\hfil\epsfxsize=5.5cm\epsfbox{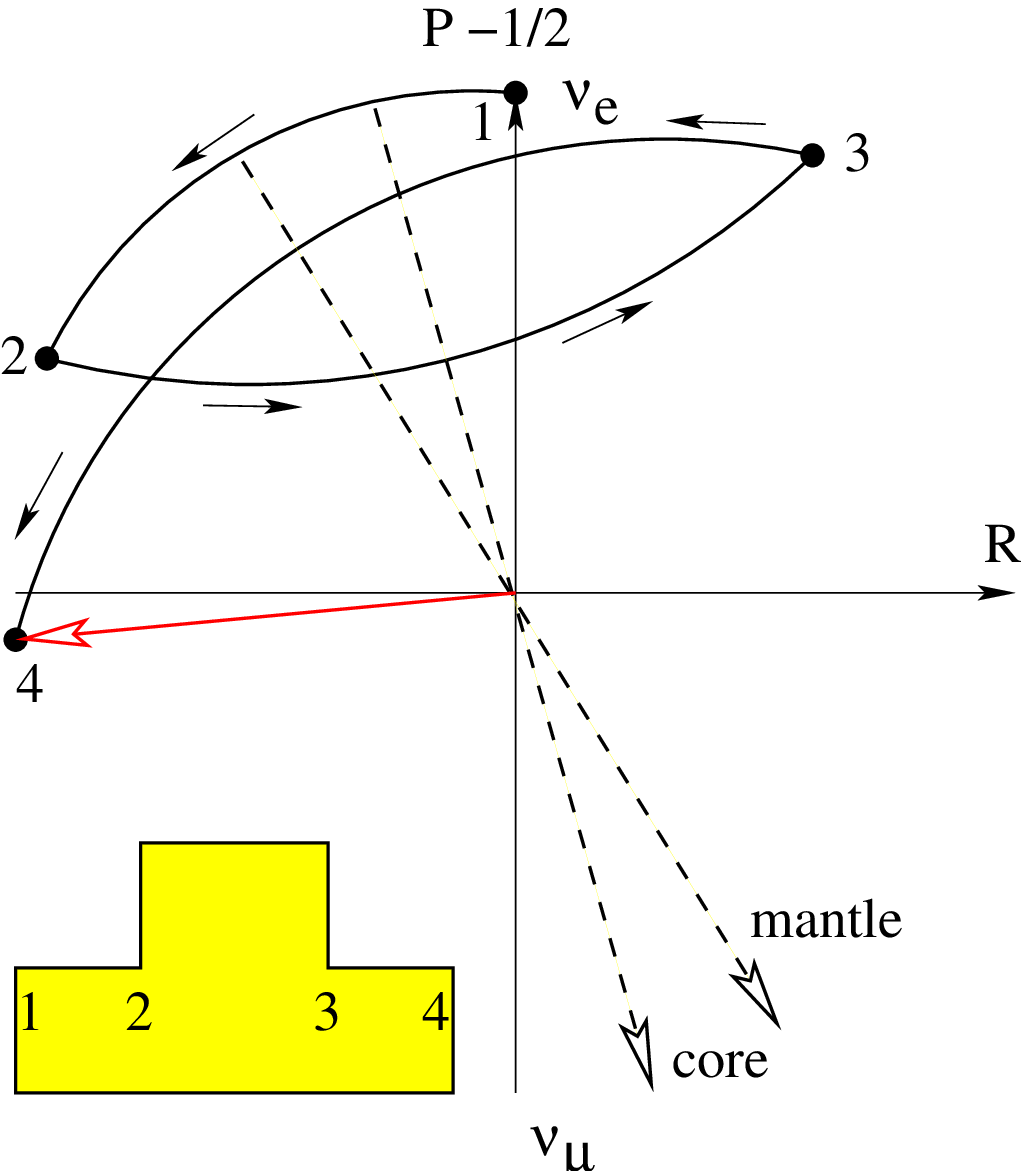}\hfil\epsfxsize=6.0cm\epsfbox{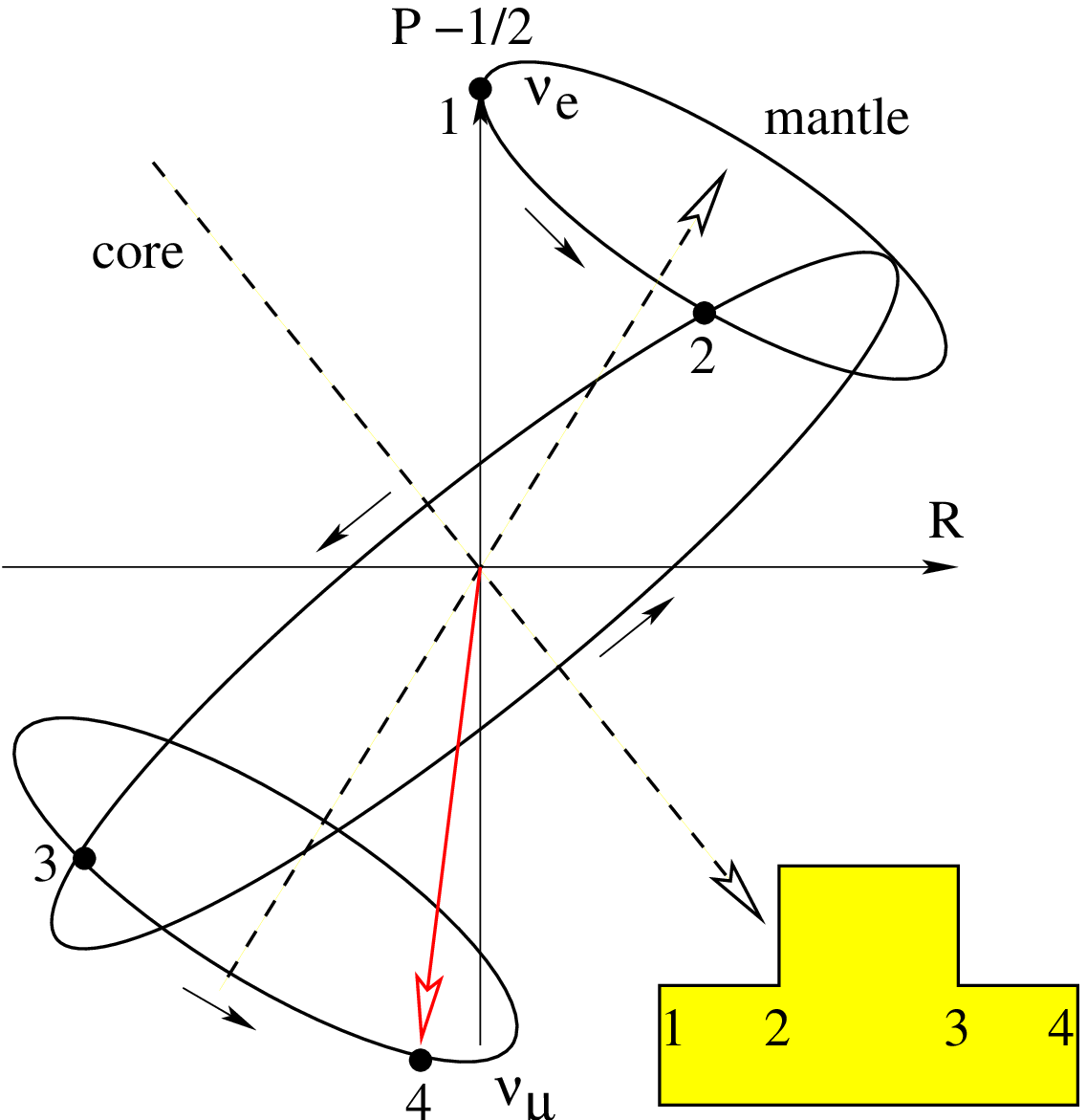}\hfil}
\caption{\small Graphic representation of the parametric enhancement 
of neutrino oscillations. The notations are explained in 
fig.~\ref{graphmsw} 
and in Eqs.~(\ref{graphnu}, \ref{axisb}).  The 
left panel: the neutrino energy  
above the resonance energy in the mantle (matter suppressed mixing).  
The right panel: the neutrino energy  is between the core and mantle  
resonance energies.  The mixing in the core is large;  
in the mantle $ \Phi_1 > \pi/4$ and in the core 
$\Phi_2 >  2\pi$. 
}
\label{param}
\end{figure}
Then under condition: 
\be
s_1 c_2 \cos 2\theta_1^m  + s_2 c_1 \cos 2\theta_2^m = 0, 
\label{parres}
\ee
where 
$s_1 \equiv \sin \Phi_i/2$, $c_1 \equiv \cos \Phi_i/2$, $(i = 1, 2)$, 
which is called the parametric resonance condition~\cite{zhenia1}, 
the flavor transition can be complete. 
Simple realization of  (\ref{parres}) is 
\be
\Phi_1 = \Phi_2 = \pi
\label{simple}
\ee
which leads to $c_1 = c_2 = 0$.  
Eq. (\ref{parres}) can be satisfied 
for neutrinos (the $\nu_{\mu} - \nu_e$ channel 
with the $\Delta m_{13}^2$ and 1-3 mixing)  
with few GeV energies which cross the core of the Earth \cite{ADLS,Pet}. 
These neutrinos propagate in  three layers  of matter: mantle-core-mantle. 
In the approximation of constant densities inside the layers, 
the profile can be considered 
as a part of the castle wall profile \cite{parliu}. 

For small mixings in both layers, the  maximal enhancement occurs when 
the condition (\ref{simple}) is satisfied (fig.\ref{param}, left panel). 
Apparently for few more  layers that would lead to maximal transition. 
On the other hand,  even three layers are enough to get nearly maximal 
transition provided that 
the mixings in matter are not small (fig.~\ref{param}, right panel).  

For $\Delta m_{13}^2 = 2 \cdot 10^{-3}$ eV$^2$, the strongest parametric  
effect (fig.~\ref{param} right panel) can be observed in a   
sample of the atmospheric neutrinos with $E > 2.5$ MeV (2 times larger 
than the energies of multi-GeV sample~\cite{ADLS}). 
Manifestation is the excess of the e-like events for the core crossing 
trajectories. It can be seen 
as an enhanced up-down asymmetry of the $e$-like events~\cite{ADLS,ruiz}.  

\subsection{Neutrino oscillations in the low density medium}  

The condition of low density,  
\be
V(x) \ll \frac{\Delta m^2}{2E}, 
\ee
means that the potential energy is much smaller than the kinetic energy.   
In this case one can use small parameter 
$\epsilon(x)$ (\ref{earth}) to develop the perturbation theory 
\cite{ara1}.

For the LMA oscillation parameters and the solar and 
supernova neutrinos:   $\epsilon(x) = (1 - 3)\cdot 10^{-2}$.  
The relevant channel of oscillations is 
the mass-to-flavor transition,  $\nu_2 \rightarrow \nu_e$, since both 
the solar and SN neutrinos arrive  at the Earth as the  incoherent fluxes 
of mass eigenstates. The probability of $\nu_2 \rightarrow \nu_e$  
can be written as 
$P_{2e} = \sin^2 \theta + f_{reg}$,  
and oscillations  appear in the first order in $\epsilon$. 
Using the $\epsilon-$  perturbation theory 
the following expression for the  regeneration factor  
$f_{reg}$ has been obtained \cite{ara1, Akhlow}
\be
f_{reg} = 
\frac{1}{2} \sin^2 2\theta \int_{x_0}^{x_f}dx V(x) \sin \Phi_m (x \rightarrow 
x_f).
\label{reg}
\ee
Here $x_0$ and $x_f$ are the initial 
and final points of propagation correspondingly,  
$\Phi_m (x \rightarrow x_f)$ is the adiabatic phase (\ref{phase}) acquired 
between a given point of trajectory, $x$, and final point, $x_f$. 
The latter feature has important consequence leading to the attenuation effect. 

In the second order in $V$, and therefore $\epsilon$,  
the regeneration factor 
becomes~\cite{ara2} 
\be
f_{reg} = 
\sin^2 2\theta \left[\sin \Phi_m (x_c \rightarrow x_f)I + \cos 2\theta~ 
I^2 
\right],   
\label{reg2}
\ee
where for the symmetric profile
\be
I =  \int_{x_c}^{x_f} dx V(x) \cos \Phi_m (x_c \rightarrow x). 
\label{inte}
\ee
In $I$ the integration proceeds from the central point of trajectory, 
$x_c$,  
to the final point, $x_f$. 
The phase is calculated from $x_c$ to a given point $x$.   
Essentially,  the integral $I$ plays the role of expansion parameter  
and  it can be estimated as $I < 2 E V_{max}/\Delta m^2 = \epsilon_{max}$.

The perturbation theory~\cite{ara2} can be  improved  
if the expansion is performed 
with respect to certain  potential $V_0$ rather than zero. 
The effective expansion parameter becomes smaller. 
 
\subsection{Attenuation effect}

Eq. (\ref{reg}) allows one to estimate  sensitivity of the oscillation 
effects to  
structures of the density profile~\cite{ara1}. Consider some structure in the point 
$x$ of the trajectory at the distance $d \equiv x_f - x$ from the detector. 
According to (\ref{reg}) for the mass-to-flavor transition  
the potential $V(x)$ 
is integrated with $\sin \Phi_m (d)$. The larger $d$, (and therefore, the larger 
$\Phi_m (d)$),  the stronger averaging effect when   
some  integration over the 
energy is performed. So, weaker sensitivity of the oscillation 
effects to remote structures of the profile should show up.  
The integration over energy with the energy resolution function 
$R(E, E')$,  
\be
\bar{f}_{reg}(E)  = \int dE' R(E, E') f_{reg}(E'), 
\ee
can be expressed  as 
\be
\bar{f}_{reg} = 
\frac{1}{2} \sin^2 2\theta \int_{x_0}^{x_f}dx V(x)  F(x_f - x)
\sin \Phi_m (x \rightarrow x_f), 
\label{regav}
\ee
where  $F(x_f - x)$ is called the {\it attenuation} factor~\cite{ara1}. 
For the box-like function $R(E, E')$ with width  (energy resolution) $\Delta E$ we 
obtain
\be
F(d) = \frac{1}{z} \sin z, ~~~~~~ z = \frac{\pi d \Delta E}{l_{\nu} E}.
\label{att}
\ee
The factor $F(0) = 1$ and it decreases with distance. This means that the 
contribution of  remote structures to the integral (\ref{regav}) is 
suppressed. The width of the first peak of $F(d)$ 
\be
l_a = l_{\nu} \frac{E}{\Delta E}  
\ee
(corresponds to $z = \pi$)  
determines the {\it attenuation length}: at $d < l_a$ the effects of 
structures  are  not suppressed. 
The better the energy  resolution, the larger the attenuation length, and 
consequently, deeper structures can be seen by the neutrino ``microscope''. 
This explains,  {\it e.g.},   why for the solar neutrinos  
the zenith angle dependence  of the  Earth mater  effect 
is flat and there is no enhancement of the  regeneration for the 
core crossing trajectories  in  spite of  2 - 3 times larger  density.  
Indeed, for the solar neutrinos with $E \sim 10$ MeV and $\Delta E/E = 
0.3$, 
we obtain $l_a = 1000$ km and therefore the contribution of the core 
is attenuated.  On the contrary, small structures  ($\sim 10$~ km) near  
the surface can produce  strong effect.    
The attenuation length increases with energy. For $E \sim 25$ MeV 
and $\Delta E/E = 0.2$, $l_a = 4000$ km, so that the core of the Earth can be 
probed by the SN neutrinos~\cite{tom}.

Another insight into phenomena can be obtained using the adiabatic perturbation 
theory which leads to~\cite{HLS}   
\be
f_{reg} = \epsilon(R) \sin^2 2\theta \sin^2 [0.5 \Phi_m (x_0 \rightarrow x_f)] + 
\sin 2\theta Re[c(x_0 \rightarrow x_f)].  
\label{fregad}
\ee 
Here $\epsilon(R)$ is the expansion parameter at the surface of the Earth and  
\be
c(x) =  
\int_{x_0}^{x} dx'\frac{d\theta_m(x')}{dx'} e^{i\Phi_m (x'\rightarrow x)}
\label{cint}
\ee
is the amplitude of transition between the eigenstates in matter   
(the adiabaticity violation effect). 
In the adiabatic case, $c(x) = 0$, the second term in 
(\ref{fregad})  is absent. 
The adiabaticity condition is broken 
at the borders of shells only. Due to sharp density change   
we have for the $j$th border: 
$d\theta_m(x)/dx|_j = \Delta V_j \delta(x - x_j)$,  where $\Delta V_j$ is the 
jump of the potential between the shells. The integration in $c(x)$ 
is trivial and  simple computations give \cite{HLS}  
\be 
f_{reg} = \frac{2E \sin^2 2\theta}{\Delta m^2} \sin \frac{\Phi_0}{2} 
\sum_{j = 0... n-1} \Delta V_j \sin \frac{\Phi_j}{2}. 
\label{sfum}
\ee    
Here $\Phi_0$ and $\Phi_j$ are the phases acquired 
along whole trajectory and on the part of the trajectory 
inside the borders $j$. This formula corresponds to  
symmetric profile with respect to the center of trajectory. 
Using (\ref{sfum}) one can easily infer the attenuation effect. 


\section{Conclusions}


The large mixing MSW conversion provides the solution of the solar neutrino 
problem: 
it leads to determination of $\Delta m_{12}^2$ and $\theta_{12}$. 
Now we have  detailed physical picture of the  conversion and its very 
precise analytical description both inside the Sun and in the Earth. 

Interesting relation emerges between $\theta_{12}$ determined in the 
solar neutrinos and the Cabibbo angle: 
$\theta_{12} + \theta_C = \pi/2$. If not accidental, it  has  
important implication for the fundamental physics. 

The small mixing MSW conversion driven by the 1-3 mixing can be realized for 
the supernova  neutrinos. Study of these neutrinos will give information 
on the  1-3 mixing and type of mass hierarchy; it opens unique possibility 
to perform monitoring of shock wave.  

A number of matter effects  can be realized for neutrinos propagating  
inside the Earth: 
(i) the resonance enhancement of oscillations;  
(ii) the parametric effects in the multi-layer medium; 
(iii) the attenuation effect for the low energy neutrinos.  
The first two effects 
can be seen in experiments with the atmospheric and accelerator neutrinos. They  
will play important role in determination of the oscillation 
parameters and establishing the type of neutrino mass hierarchy. 
The attenuation effect  is realized for  
the solar and supernova neutrinos, it describes the loss of sensitivity to  
remote structures of the density profile. 
The effect is crucial  for the oscillation tomography of the Earth.


\begin{thebibliography}{99}
\bibitem{w1} L. Wolfenstein, Phys. Rev. D{\bf 17}, 2369 (1978); 
in {\it ``Neutrino -78"}, Purdue Univ. C3, (1978), Phys. Rev. D{\bf 20}, 
2634 (1979).

\bibitem{pontosc} B. Pontecorvo, Zh. Eksp. Theor. Fiz. {\bf 33} (1957); 
{\it ibidem}  {\bf 34}, 247 (1958). 

\bibitem{mns} Z. Maki, M. Nakagawa and S. Sakata, Prog. Theor. Phys. {\bf 28} (1962) 870. 

\bibitem{pontprob} B. Pontecorvo, ZETF, {\bf 53}, 1771 (1967) [Sov. Phys. 
JETP, {\bf 26}, 984 (1968)]; 
V. N. Gribov and B. Pontecorvo, Phys. Lett. {\bf 28B},  493 
(1969). 

\bibitem{bar} V. Barger, K. Whisnant, S. Pakvasa and  R.J. N. Phillips, 
Phys. Rev. D{\bf 22}, 2718 (1980); S. Pakvasa, in DUMAND-80, vol. {\bf 2}, 
457 (1981).

\bibitem{ms1} S. P. Mikheyev and A. Yu. Smirnov, Sov. J. Nucl. Phys.
{\bf 42}, 913 (1985),  Nuovo Cim. {\bf C9}, 17 (1986); 
S.P. Mikheev and  A.Yu. Smirnov, Sov. Phys. JETP {\bf 64}, 4  (1986).


\bibitem{paul} P. Langacker, J. P. Leville and J. Sheiman, Phys. Rev. D {\bf 27} 
1228  (1983); V. B. Semikoz, Sov. J. Nucl. Phys. {\bf 46}, 946 (1987).   

\bibitem{bet} H. Bethe,  Phys. Rev. Lett. {\bf 56}, 1305 (1986). 


\bibitem{mess} A. Messiah, Proc. of the 6th Moriond Workshop on Massive 
Neutrinos in Particle Physics and Astrophysics, eds O. Fackler and J. Tran 
Thanh Van, Tignes, France, Jan. 1986, p. 373;  
S. P. Mikheev and A.Y. Smirnov,  Sov. Phys. JETP {\bf 65},  230 (1987). 

\bibitem{ssb} A. Yu. Smirnov, D. N. Spergel, 
J. N. Bahcall, Phys. Rev. D{\bf 49} 1389 (1994).

\bibitem{lisi} E. Lisi, A. Marrone, D. Montanino, A. Palazzo and S.T. 
Petcov,  Phys. Rev. D{\bf 63} 093002 (2001); 
A. Friedland, Phys. Rev. {\bf D64}, 013008 (2001), 
and hep-ph/0106042. 

\bibitem{hax} W. C. Haxton, Phys. Rev. Lett. {\bf 57}, 1271 (1986);  
S. J. Parke, Phys. Rev. Lett. {\bf 57}, 1275 (1986);
S. P. Rosen and J. M. Gelb, Phys. Rev. D {\bf 34}, 969 (1986).




\bibitem{ms4} S. P. Mikheyev and A. Yu. Smirnov, Proc. of the 6th Moriond
Workshop on massive Neutrinos in Astrophysics and Particle Physics,
Tignes, Savoie, France  Jan. 1986  (eds. O. Fackler and J. Tran Thanh Van)
p. 355 (1986); J. Bouchez {\it et al}, Z. Phys. C {\bf 32} (1986) 499;
V. K. Ermilova, V. A. Tsarev, V. A. Chechin,
JETP Lett. {\bf 43},  453 (1986).



\bibitem{minw}
C. Lunardini, A. Yu. Smirnov, 
Nucl. Phys. B{\bf 583}, 260 (2000). 



\bibitem{MD} A. McDonald, these proceedings;  Y. Suzuki, these proceedings.



\bibitem{kam} A. Suzuki, these proceedings,  
KamLAND Collaboration, K. Eguchi et al., Phys. Rev. Lett., {\bf 90},
 021802 (2003);  T. Araki et al.,  hep-ex/0406035. 




\bibitem{HSsol}
P. C. de Holanda, A.Yu. Smirnov, Astropart. Phys. {\bf 21}, 287 (2004). 

\bibitem{john}J. N. Bahcall, M.H. Pinsonneault,  Phys. Rev. Lett. {\bf 92}, 121301 
(2004). 


\bibitem{lisiev}G. Fogli, E. Lisi, New J. Phys. {\bf 6}, 139 (2004);  
G.L. Fogli, E. Lisi, A. Marrone, A Palazzo, 
Phys. Lett. B{\bf 583}, 149 (2004).  

\bibitem{HLS}P. C. de Holanda, Wei Liao, A. Yu. Smirnov,  
Nucl. Phys. B{\bf 702}, 307 (2004). 


\bibitem{DS}
A. S. Dighe, A. Yu. Smirnov, Phys. Rev. D{\bf 62},  033007 (2000);   
C. Lunardini,  A. Yu. Smirnov,  JCAP {\bf 0306}, 009 (2003). 

\bibitem{nun}
H. Minakata,  H. Nunokawa, Phys. Lett. B {\bf 504},  301 (2001). 

\bibitem{barg}
V. Barger,  D. Marfatia, B.P. Wood, Phys. Lett. B {\bf 532}, 19 (2002). 

\bibitem{snsato}
K. Takahashi, K. Sato, A.  Burrows,  
T. A. Thompson, Phys. Rev. D {\bf 68},  113009 (2003). 

\bibitem{raffelt} G. Raffelt, these proceedings. 




\bibitem{SF} R.C. Schirato and G. M. Fuller, astro-ph/0205390.

\bibitem{Tak}
K. Takahashi, K. Sato,  H. E. Dalhed, J.R. Wilson, 
Astropart. Phys. {\bf 20},  189 (2003).


\bibitem{lisis} G.L. Fogli, E. Lisi, D. Montanino, A. Mirizzi,
Phys. Rev. D {\bf 68}, 033005 (2003). 


\bibitem{munich}
R. Tomas,  {\it et al.}, astro-ph/0407132. 

\bibitem{LS04}
C. Lunardini, A. Yu. Smirnov, Phys. Rev. D {\bf 63}, 073009 (2001);   Astropart. 
Phys. {\bf 21}, 703 (2004);   
M. L. Costantini, A. Ianni, F. Vissani, Phys. Rev. D {\bf 70}, 043006 (2004). 

\bibitem{lind}
M. Freund, T. Ohlsson, Mod. Phys. Lett. A {\bf 15}, 867 (2000);  
M. Freund, M. Lindner, S.T. Petcov, A. Romanino, 
Nucl. Phys. B {\bf 578}, 27 (2000);   
T. Ohlsson, H. Snellman, Phys. Lett. B {\bf 474}, 153 (2000).    


\bibitem{ADLS} E. K. Akhmedov, A. Dighe, P. Lipari, A.Y. Smirnov,  
Nucl. Phys. B {\bf 542}, 3 (1999).

\bibitem{ruiz} 
J. Bernabeu, S. Palomares-Ruiz, A. Perez, S.T. Petcov, 
Phys. Lett. B {\bf 531}, 90 (2002), 
J. Bernabeu, S. Palomares Ruiz, S.T. Petcov, Nucl. Phys. B {\bf 
669}, 255 (2003).




\bibitem{ETC} V. K. Ermilova, V. A. Tsarev and V. A. Chechin, Kr. Soob,
Fiz. [Short Notices of the Lebedev Institute] {\bf 5}, 26 (1986).

\bibitem{Akh} E. Kh. Akhmedov, 
Yad. Fiz. {\bf 47}, 475 (1988) [Sov. J. Nucl. Phys. {\bf 47}, 301 (1988)].



\bibitem{parliu}
Q. Y. Liu, A. Yu. Smirnov,  Nucl. Phys. B {\bf 524}, 505 (1998); 
Q. Y. Liu, S. P. Mikheyev, A. Yu. Smirnov, Phys. Lett. B{\bf 440}, 319 
(1998). 

\bibitem{Pet} S. T. Petcov, Phys. Lett. {\bf B434} (1998) 321; 
M. Chizhov, M. Maris, S.T. Petcov,  hep-ph/9810501;
M.V. Chizhov, S.T. Petcov, Phys. Rev. Lett. {\bf 83}, 1096 (1999);
Phys. Rev. D {\bf 63}, 073003 (2001).  



\bibitem{zhenia1}
E.K. Akhmedov, Nucl. Phys. B {\bf 538},  25 (1999), [hep-ph/9805272];  
hep-ph/9903302; 
Pramana  {\bf 54}, 47 (2000), [hep-ph/9907435].  
E.K. Akhmedov, A.Yu. Smirnov,
Phys. Rev. Lett. {\bf 85}, 3978 (2000), and  hep-ph/9910433.








\bibitem{ara1}A. N. Ioannisian, A.Yu. Smirnov,
Phys. Rev. Lett. {\bf 93}, 241801 (2004). 


\bibitem{Akhlow}
E. Kh. Akhmedov, M.A. Tortola, J.W.F. Valle, JHEP {\bf 0405}, 057 (2004).  

\bibitem{ara2} A. N. Ioannisian, N.A. Kazarian, A.Yu. Smirnov, D. Wyler, 
hep-ph/0407138. 

\bibitem{tom}
M. Lindner, T. Ohlsson, R. Tomas, W. Winter, Astropart. Phys. {\bf 19}, 755 
(2003). 

\end{thebibliography}
\end{document}